\documentclass[prx,letterpaper,nobalancelastpage,twocolumn,superscriptaddress,nofootinbib]{revtex4-2}

\usepackage{graphicx}
\usepackage{amsmath}
\usepackage{bbold}
\usepackage{amssymb}
\usepackage[english]{babel}
\usepackage{color}
\usepackage[version=4]{mhchem}
\usepackage[hidelinks]{hyperref}
\usepackage{dsfont}
\usepackage{placeins}
\usepackage{mathtools}
\usepackage[normalem]{ulem}
\usepackage{lipsum}

\setcitestyle{super}

\usepackage{soul}
\newcommand{\ket}[1]{| #1 \rangle}
\newcommand{\bra}[1]{\langle #1 |}

\newcommand{\Caltech}{California Institute of Technology, Pasadena, CA 91125, USA}

\newcommand{\Stanford}{Department of Electrical Engineering, Stanford University, Stanford, CA, USA}

\let\oldchi\chi
\renewcommand{\chi}{%
  \raisebox{0.44ex}{$\oldchi$}%
}
\usepackage{caption}

\DeclareCaptionLabelSeparator{bar}{ \textbf{\textbar}~}
\usepackage{enumitem}
\setlist{nolistsep}
\begin{document}

\title{Erasure-cooling, control, and hyper-entanglement of motion in optical tweezers}
\author{Adam L. Shaw,$^{1,\ast,\dag}$ Pascal Scholl,$^{1,\ast,\ddag}$ Ran Finkelstein,$^{1,\ast,\mathparagraph}$ Richard Bing-Shiun Tsai,$^1$ \\Joonhee Choi,$^{1,2}$ and Manuel Endres$^{1,\mathsection}$\\}
\noaffiliation
\affiliation{\Caltech}
\affiliation{\Stanford}

\begin{abstract}
Coherently controlling the motion of single atoms in optical tweezers would enable new applications in quantum information science. To demonstrate this, we first prepare atoms in their motional ground state using a species-agnostic cooling mechanism that converts motional excitations into erasures -- errors with a known location. This cooling mechanism fundamentally outperforms idealized traditional sideband cooling, which we experimentally demonstrate. By coherently manipulating the resultant pure motional state, we perform mid-circuit readout and mid-circuit erasure detection via local shelving into motional superposition states. We finally entangle the motion of two atoms in separate tweezers and generate hyper-entanglement by preparing a simultaneous Bell state of motional and optical qubits, unlocking a large new class of quantum operations with neutral atoms. 

\end{abstract}
\maketitle

The ability to store, carry, and manipulate quantum information is essential to the success of quantum science applications~\cite{Shor1995,Knill1996,Toth2014,Wehner2018,Feihu2020}.
Tremendous effort has been invested in a wide array of quantum platforms to develop efficient quantum information carriers~\cite{Wendin2017,Bruzewicz2019,Slussarenko2019,Shandilya2022,Kang2023}.
In particular, for neutral atom based platforms, quantum information has so far been encoded in electronic or nuclear states~\cite{Saffman2016,Levine2019,Browaeys2020,Ma2023}, the manipulation of which is often limited by atomic thermal motion~\cite{Leseleuc2018,Schine2022,Bluvstein2022,Shaw2023B}. \footnote{\label{a}$^*$ These authors contributed equally to this work.}
\footnote{\label{b}$^{\dag}$ Current affiliation: Department of Physics, Stanford University}
\footnote{\label{c}$^{\ddag}$ Current affiliation: PASQAL, 7 rue Léonard de Vinci, 91300 Massy, France}
\footnote{\label{c}$^{\mathparagraph}$ Current affiliation: School of Physics and Astronomy, Tel Aviv University, 69978, Israel}
\footnote{\label{d}$^{\mathsection}$ mendres@caltech.edu} 

\begin{figure}[t!]
	\centering
	\includegraphics[width=\columnwidth]{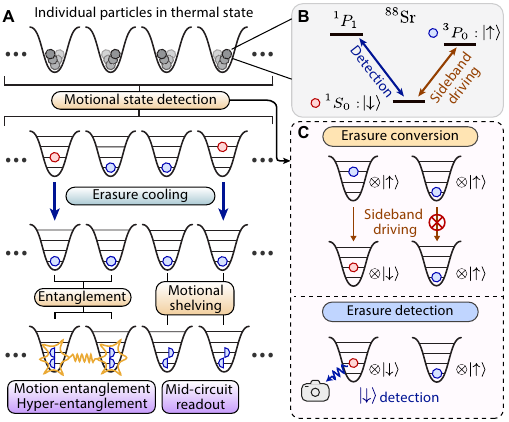}
	\caption{\textbf{Scheme for erasure cooling, coherent manipulation, and entanglement of motional states.} 
 \textbf{(A)} Motional degrees of freedom in optical tweezers can be used as quantum information carriers. We first initialize the atoms in their motional ground state using a cooling mechanism based on motional state error-correction. With such a low-entropy state at hand, we subsequently demonstrate (i) the use of a coherent superposition of motional states to perform mid-circuit readout of an optical qubit, and (ii) motion-entanglement as well as hyper-entanglement: simultaneous spin-entanglement and motion-entanglement. \textbf{(B)} Relevant pseudo-spin levels, and detection transition used in this work. \textbf{(C)} Protocol to detect the atom's motional state. We first convert motional excitation errors into erasure errors, i.e., errors with a known location~\cite{Grassl1997, Wu2022}. We accomplish this by selectively transferring motion-excited atoms to the electronic ground state, ${}^1S_0$ (red), whilst motional ground state atoms are shelved in $^3P_0$ (blue). We then selectively detect atoms in ${}^1S_0$, which does not perturb atoms in $^3P_0$.  
 }
	\vspace{0.5cm}
	\label{Fig1}
\end{figure}

However, this same motion can in principle be used to encode and store quantum information. The motion of atoms trapped in nearly-harmonic potentials, such as optical tweezers, is quantized~\cite{Brown2023,Hartke2022}, and acts essentially as a bosonic degree of freedom, similar to auxiliary bosonic states in ion traps~\cite{Knight2003,Fluhmann2019} and superconducting circuits~\cite{Heeres2017,Campagne-Ibarcq2020}. Recent experiments exploiting this resource have demonstrated squeezing of motional states~\cite{Brown2023} and creation of a quantum register of fermion pairs~\cite{Hartke2022}. Unlike electronic or nuclear states, motional states do not interact directly with electromagnetic fields and are thus robust to environmental effects~\cite{Singh2023}. Further, their bosonic nature may be used to implement quantum error correction schemes with bosonic degrees of freedom such as Gottesman-Kitaev-Preskill (GKP) codes~\cite{Gottesman2001,Fluhmann2019}, or as a resource for studying lattice gauge theories~\cite{Jordan2014,Macridin2022}. As motional states are independent from electronic or nuclear states, information could be encoded in multiple degrees of freedom -- for instance, motional and electronic -- independently. This would enable multiple bits of quantum information to be encoded in a single atom, and atoms could be entangled through two or more degrees of freedom to create so-called hyper-entanglement~\cite{Ren2014, Deng2017,Shi2021,Hu2021,Hu2018,Graham2015}, a resource which has so far only been demonstrated for photonic qubits. However, to realize these prospects first requires an efficient cooling scheme to prepare atoms in their motional ground state with high fidelity. Here we demonstrate such a technique, and use the resultant states as a resource to show the potential of motional states in enabling key tasks in quantum information processing (Fig.~\ref{Fig1}A).

Our cooling mechanism is species-agnostic and relies on (i) conversion of motional excitations into detectable, localized errors~\cite{Lee2023}, also called erasures~\cite{Grassl1997, Wu2022}, and (ii) their active correction. Similar techniques have recently been applied to quantum simulation and computing tasks with nuclear or optical transitions~\cite{Scholl2023,Ma2023} of atoms in optical tweezers. We use the resultant arrays of ground state atoms to demonstrate coherent transduction from electronic to motional superposition states, mid-circuit readout~\cite{Singh2023,Deist2022,Graham2023,Lis2023,Norcia2023} of an optical transition, and mid-circuit erasure detection~\cite{Ma2023}. We then extend beyond the single-particle regime by entangling the motion of two atoms in separate tweezers through Rydberg interactions~\cite{Madjarov2020,Levine2019}, while leaving their electronic degrees of freedom separable. Finally, we generate hyper-entanglement~\cite{Ren2014, Deng2017} in our matter-based qubits by creating a Bell state both in the motional and electronic degrees of freedom simultaneously.

\vspace{1mm}
\noindent\textbf{Erasure correction cooling}\newline
Our cooling scheme, which we term \textit{erasure correction cooling} (ECC), is based on detecting atoms in motion-excited states -- while motion-ground state atoms are protected through shelving into an ancillary state -- and then either removing or selectively re-cooling such motional excitations. This is similar to, for instance, atomic rearrangement techniques~\cite{Endres2016,Barredo2016} which detect and correct errors in trap loading in atom arrays via fluorescence imaging followed by dynamic trap movement, without discarding atoms in already occupied traps. We broadly consider such rearrangement techniques and our erasure cooling technique as forms of \textit{classical} erasure conversion. This is in contrast to \textit{quantum} erasure conversion where an additional internal degree of freedom is left unprojected during the measurement, for instance as has been demonstrated for atom arrays using a nuclear spin qubit~\cite{Ma2023}, a Rydberg qubit~\cite{Scholl2023}, and which later in this manuscript we will demonstrate for a qubit defined via low-lying states in the motional manifold. Erasure conversion here refers to the general process of performing projective measurement on an array of atoms in such a way that we raise a flag on erroneous atoms while not perturbing observables of interest for error-free atoms.

Our experiment~\cite{Madjarov2020,SM} is based on trapping individual strontium atoms in arrays of optical tweezers~\cite{Cooper2018,Norcia2018,SM} (Fig.~\ref{Fig1}A and B). Atoms are initialized in their ground electronic state $^1S_0$ and initially cooled with Sisyphus cooling to an initial motional ground state population of $P_0' = 0.77^{+1}_{-1}$. To perform ECC, we convert remaining motional excitations into erasure errors~\cite{Grassl1997, Wu2022} (Fig.~\ref{Fig1}C) by working in the sideband resolved regime on the magic-trapped transition between $^1S_0$ and the metastable excited state $^3P_0$.

In what follows, we consider a single motional degree of freedom, which is defined along the $^1S_0\leftrightarrow{^3P_0}$ laser beam propagation axis. Results in the main text focus on the tightly-confined, radial direction of the optical tweezers; we find similar results along the weakly-confining, axial direction~\cite{SM}. We use numerical notation for motional states (e.g. $\ket{0}$ for the motional ground state), and pseudo-spin notation for electronic states (e.g. $\ket{{\downarrow}}$ is $^1S_0$ and $\ket{{\uparrow}}$ is $^3P_0$); for the latter we will use the terms spin, electronic state, and optical qubit interchangeably. We further denote their tensor product as e.g. \mbox{$\ket{0} \otimes \ket{{\uparrow}} = \ket{{0,\uparrow}}$}.

\begin{figure*}[t!]
	\centering
	\includegraphics[width=\textwidth]{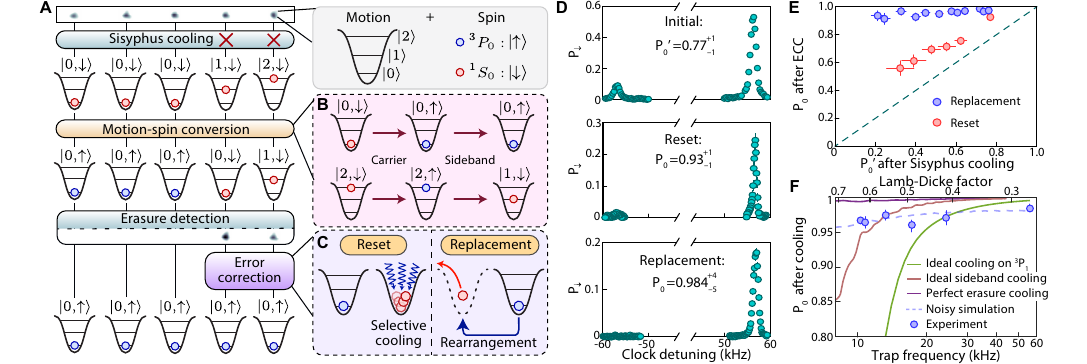}
	\caption{\textbf{Improving motional ground state preparation with erasure correction cooling.} \textbf{(A,B)} After cooling the atoms with Sisyphus cooling, we apply spin-motion conversion, which leaves atoms in a different pseudo-spin state depending on whether they do or do not have motional excitations, as shown explicitly for the initial states $\ket{2,{\downarrow}}$ and $\ket{0,{\downarrow}}$ in \textbf{(B)}. We then detect atoms with motional excitations by imaging the pseudo-spin state $\ket{{\downarrow}}$, without perturbing atoms in the motional ground state, which reside in $\ket{{\uparrow}}$. \textbf{(C)} We correct tweezers containing atoms with motional excitations in two different ways: (i) by reapplying Sisyphus cooling selectively on such atoms (reset), or (ii) by replacing such atoms using rearrangement (replacement). \textbf{(D)} Probability, $P_0$, for an atom to be in the motional ground state, $\ket{0}$, measured via sideband spectroscopy. We show the results without erasure correction cooling, performing erasure correction with reset, and erasure correction with replacement. \textbf{(E)} Post-correction motion-ground state probability, $P_0$, as a function of intentionally varied motion-ground state probability after Sisyphus cooling, $P_0'$. We show the results for both replacement (blue markers) and reset (red markers). The replacement method is largely insensitive to the Sisyphus cooling efficiency. The green dashed line is a guide to the eye which represents unity gain. \textbf{(F)} $P_0$ as a function of the trap frequency $\omega$. Our experimental erasure cooling results (blue markers) using the replacement method are in good agreement with an \textit{ab initio} error model including known imperfections (blue dashed line). We compare against two variants of sideband cooling, one using the  $^3P_1$ state of strontium (green line), and one which is an idealization representing the ultimate limit of sideband cooling for given trap and Rabi frequencies, and Lamb-Dicke factor~\cite{SM} (red line). Our experimental erasure cooling outperforms both sideband cooling variants in the regime of low trap frequency; an error-free simulation of erasure cooling is superior to sideband cooling for all frequencies. Error bars represent the standard error of the mean.
 }
	\vspace{0.5cm}
	\label{Fig2}
\end{figure*}

Erasure correction cooling proceeds with the following steps (Fig.~\ref{Fig2}A and B): first we drive all atoms into the excited electronic state using the motion-preserving \textit{carrier} transition, \mbox{$\ket{n,\downarrow}\rightarrow\ket{n,\uparrow}$}, where $n$ is an arbitrary motional level. Second, we perform a spin-motion coupling \textit{sideband} transition, $\ket{n,\uparrow}\leftrightarrow\ket{n-1,\downarrow}$, ideally leaving only motional ground state atoms in the excited electronic state. We then perform fluorescence imaging of the ground electronic state, while leaving atoms in $\ket{0,\uparrow}$ unperturbed~\cite{Scholl2023}. Thus we directly image the location of motion-excited atoms, for which we can then perform error correction.

This correction can be done in two ways (Fig.~\ref{Fig2}C), either: (1) \textit{Reset}, where we reapply Sisyphus cooling with site-selectivity~\cite{SM}, or (2) \textit{Replacement}, by discarding hot atoms and replacing them with cold atoms using atom rearrangement~\cite{Endres2016,Barredo2016}. Following ECC, we measure $P_0$ via sideband spectroscopy on the $\ket{{\downarrow}} \rightarrow \ket{{ \uparrow}}$ transition (Fig.~\ref{Fig2}D). Starting from $P_0' = 0.77^{+1}_{-1}$ after Sisyphus cooling, we obtain a final motional ground state population $P_0 = 0.93^{+1}_{-1}$ with one round of reset, and $P_0 = 0.984^{+4}_{-5}$ with replacement; for the latter, if we post-select on erasure detection, rather than applying active correction, we achieve $P_0 = 0.995^{+3}_{-4}$. The maximum $P_0$ after reset is limited by doing only one round of cooling, with a theoretical maximum of \mbox{$P_0=P_0'+(1-P_0')P_0'\approx0.95$}. In Fig.~\ref{Fig2}E we show the efficacy of ECC as a function of $P_0'$ obtained after Sisyphus cooling (allowing up to 3 rounds of reset cooling for the lowest $P_0'$ values); in general we find that ECC with replacement is insensitive to the initial cooling temperature.

Further, ECC outperforms even idealized, noise-free, conventional laser cooling in certain realistic parameter regimes (Fig.~\ref{Fig2}F). We compare noise-free simulations~\cite{SM} of ECC (purple line) against noise-free simulations of cooling on the conventional $^1S_0 \rightarrow {}^3P_1$ intercombination line~\cite{Cooper2018, Norcia2018} (green line), and of idealized sideband cooling performed with infinitely small linewidth (red line), but including an effective spontaneous decay in order to remove entropy from the system~\cite{Zhang2022,SM}. We find idealized ECC always outperforms idealized sideband cooling, but particularly so in the limit of low trap depth, where the sideband-resolved condition breaks down~\cite{SM}. This is so even when including realistic noise sources in the ECC simulation (dashed line), and for experimental results as well (blue points). Our results demonstrate that the high-fidelity cooling can be achieved practically at a much lower trap depth than normally expected, potentially opening the door to greatly expanding the size of tweezer arrays.

\vspace{1mm}
\noindent\textbf{Preparing motional superposition states}\newline
Having obtained motional ground state atoms with high fidelity thanks to erasure correction cooling, we detail how we make use of this low-entropy atom array for quantum information purposes. We focus on controlling only a single motional degree of freedom, and in particular focus on the lowest two motional levels, which we refer to as the \textit{motional qubit}. The isolation of these two motional states does not rely on trap anharmonicity and is instead achieved via driving sideband transitions which only couple the motional ground state to the first excited state, therefore leaving higher energy levels unpopulated. Extensions of our scheme and results are readily generalized to control over all three axes, and to control over higher energy states in the motional manifold, which may enable various applications~\cite{Fluhmann2019}. We will use a trap frequency (equivalently, motional qubit energy spacing) of $\omega/(2\pi) \simeq 35.5 \, \text{kHz}$.

\begin{figure*}[t!]
	\centering
	\includegraphics[width=\textwidth]{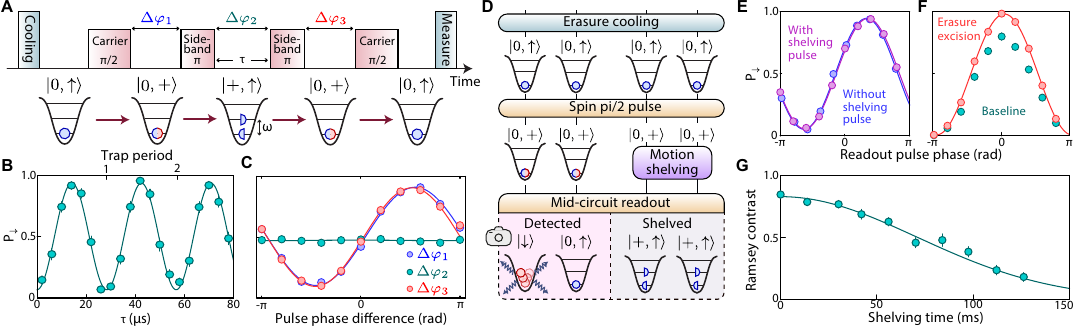}
	\caption{\textbf{Spin-motion transduction and mid-circuit readout.} \textbf{(A)} Protocol used to transfer a coherent superposition of pseudo-spin states into a superposition of motional states. After cooling the atoms and preparing the spin-superposition state $\ket{0,+}$, we apply a sideband pulse which drives the $\ket{0,\downarrow} \rightarrow \ket{1, \uparrow}$ transition, allowing us to obtain the motion-superposition state $\ket{+,\uparrow}$. We then revert the process in order to read out the coherence of the created state. \textbf{(B)} Probability to obtain the atom in $\ket{{\downarrow}}$ as a function of holding time in $\ket{+,\uparrow}$. We observe an oscillation with frequency $35.3^{+0.3}_{-0.3} \, \text{kHz}$, consistent with our trapping frequency $\omega/(2\pi) \simeq 35.5 \, \text{kHz}$. \textbf{(C)} Probability to obtain the atom in $\ket{{\downarrow}}$ as a function of the phase difference between the pulses in \textbf{(A)}. The atom is sensitive to the laser phase while in a pseudo-spin superposition state (red and blue). However, when in the motion-superposition state $\ket{+,\uparrow}$, the atom is not sensitive to the laser phase (with a contrast consistent with 0), and is hence insensitive to laser phase noise (green). \textbf{(D)} Local pseudo-spin coherence preservation using motional state shelving. We prepare the $\ket{0,+}$ state and locally shelve some atoms into $\ket{+,\uparrow}$ by lowering their trap depth~\cite{Shaw2023A} and applying site-selective sideband drive. We detect atoms in $\ket{{\downarrow}}$ to project the unshelved atoms into $\ket{{\downarrow}}$ or $\ket{{\uparrow}}$. The shelved atoms are not affected by the detection, hence the coherence is preserved within the motional states. \textbf{(E)} The shelving operation does not significantly perturb the unshelved atoms, which we test by performing Ramsey spectroscopy on the unshelved atoms either with (purple) or without (blue) the shelving operation in the dark time, which each yield oscillations with consistent contrasts of $0.89^{+1}_{-1}$. \textbf{(F)} Obtained pseudo-spin coherence for the shelved atoms after the sequence described in \textbf{(D)}. We obtain a bare one-way transduction fidelity of $0.943_{-3}^{+3}$, further increased to $0.993_{-3}^{+3}$ with erasure detection and excision~\cite{SM}. \textbf{(G)} Motional coherence in an echo sequence~\cite{SM} as a function of shelving time. We obtain a typical $1/e$ coherence time of $\sim 100 \, \text{ms}$, which is long compared to our fast detection time of $24 \, \mu \text{s}$. Error bars represent the standard error of the mean; points represent data and lines represent fits.
 }
	\vspace{0.5cm}
	\label{Fig3}
\end{figure*}

We first show the preparation and readout of a coherent superposition state of the motional qubit (Fig.~\ref{Fig3}A). Having isolated atoms in $\ket{0,\uparrow}$ with ECC, we prepare a spin-superposition state, and transduct this superposition onto the motional qubit by performing spin-motion coupling via sideband driving. This yields the state $\ket{+,0} = \frac{1}{\sqrt{2}}(\ket{{0}}+\ket{{1}})\otimes\ket{{\uparrow}}$. We hold this state unperturbed to evolve coherently, and then read its state by reversing the preparation sequence. The resultant signal (the probability $P_{\downarrow}$ for the atom be in $\ket{{\downarrow}}$) after the entire sequence shows Ramsey oscillations at a frequency of  $35.3^{+3}_{-3} \, \text{kHz}$, consistent with the programmed trap frequency, as is expected for the motional qubit (Fig.~\ref{Fig3}B).

After the transduction, the motional qubit is sensitive to dephasing largely only from noise on the trapping laser \textit{power} (which controls the relative qubit energy spacing). This is in contrast to optical qubits, which may dephase due to noise on the addressing laser \textit{frequency}, often limiting the coherence time for optical qubits based on ultranarrow transitions~\cite{Madjarov2019}. We demonstrate this insensitivity by fixing the hold time and varying the laser phase between each preparation or readout pulse (Fig.~\ref{Fig3}C). $P_{\downarrow}$ is insensitive to laser phase variation between the two sideband pulses (green points) while the atom is in a motional superposition state, showing a fitted contrast of $0.002^{+6}_{-6}$, consistent with zero (green line). By contrast, while the atom is in an optical superposition state (red and blue points and lines), much higher sensitivity is observed; importantly, in practice the time between pulses is short enough to avoid significant dephasing. Thus, our results show that motional levels can be used to store information in a subspace that is insensitive to phase noise~\cite{Leseleuc2018} on the laser which drives the optical qubit transition. Other extrinsic noise sources, such as environmental magnetic and electric field noise should behave similarly, assuming they minimally affect the trap frequency. This overall insensitivity makes the motional qubit a promising storage subspace for quantum information applications.

\vspace{1mm}
\noindent\textbf{Mid-circuit readout with motional state shelving}\newline
Having demonstrated our capability to prepare motion-superposition states, we now show that this can be used to perform mid-circuit readout~\cite{Singh2023,Deist2022,Graham2023,Lis2023,Norcia2023} (Fig.~\ref{Fig3}D), which is a necessary step in error correcting codes~\cite{Terhal2015}, and which enables various near term applications such as measurement-based generation of long-range entangled states~\cite{Lee2022,Verresen2021} and enhanced metrology~\cite{Rosenband2013}.

To do so, we first divide the atom array into two ensembles $A$ and $B$, and prepare both in the motion-superposition $\ket{0,+}$ state. We then selectively lower the trap depths in $B$, while holding $A$ constant~\cite{Shaw2023A}. This induces a reduction in trap frequency for atoms in $B$, and thus changes the motional qubit energy spacing. We can then selectively apply sideband driving only to the ensemble $B$, hence transferring only atoms in $B$ into the $\ket{+,\uparrow}$ state~\cite{Shaw2023A}, leaving $A$ largely unperturbed. To test that $A$ is not perturbed, we perform a readout pulse with varying phase either with or without the shelving operation on $B$. Both cases show the same Ramsey contrast in $A$ of $0.89^{+1}_{-1}$, with a uncontrolled phase shift of only $0.018^{+9}_{-9}\pi$ which could likely be further reduced through more careful calibration (Fig.~\ref{Fig3}E).

Having shelved atoms in $B$, we apply a global, fast detection of atoms in the electronic ground state~\cite{Scholl2023}, which projects the spin state of atoms in $A$, while preserving the coherence of atoms in $B$. We finally unshelve atoms in $B$, and read out their coherence (Fig.~\ref{Fig3}F). The shelving and unshelving procedure is designed to apply a motional echo composed of four sideband pulses in order to eliminate shot-to-shot fluctuations of the trap depth~\cite{SM}. We obtain a bare contrast of $0.79_{-1}^{+1}$ after performing these four sideband pulses, which corresponds to a state preparation and measurement (SPAM) corrected~\cite{SM} one-way, optical-to-motion transduction fidelity of $0.943_{-3}^{+3}$. 

Further, if the transduction fails, then atoms in $B$ are detected as erasures during the fast image, which we then excise to obtain an overall contrast of $0.972_{-10}^{+10}$ which corresponds to a SPAM-corrected one-way transduction fidelity of $0.993_{-3}^{+3}$. One limitation to this number arises from trap frequency inhomogeneity of a few percent across the array, which could be improved with better uniformity of the trap waists (and depths), or by employing more advanced pulse sequences designed to be insensitive to such inhomogeneities~\cite{Gong2023}. Further limitations may arise from time-dependent variations of the frequency of the addressing laser, or the trap frequency, which for instance is already a limitation to driving of the carrier transition. Since the fast image has negligible impact on the coherence of motional states in $B$, this protocol demonstrates \textit{quantum} erasure conversion, where the protected internal degree of freedom is the motional qubit, analogous to for instance the nuclear spin qubit or Rydberg qubit utilized in demonstrations of quantum erasure conversion with Yb~\cite{Ma2023} or Sr~\cite{Scholl2023} atom arrays, respectively. Further, we perform this erasure conversion in a mid-circuit fashion (as the measurement is much faster than the coherence time of the motional qubit), which has only previously been demonstrated for neutral atom systems with Yb atoms~\cite{Ma2023}.

We measure the coherence time of the motional states for atoms in $B$ (Fig.~\ref{Fig3}G) using a motion-echo sequence~\cite{SM}, and obtain a typical lifetime of ${\sim}100 \, \text{ms}$, which is long compared to our imaging time of $24\, \mu \text{s}$ using our fast detection scheme~\cite{Scholl2023}. This coherence time is likely improvable through better stabilization and limiting tweezer-induced Rayleigh scattering~\cite{SM}. 

In contrast to other schemes~\cite{Singh2023,Deist2022,Graham2023,Lis2023,Norcia2023}, our method does not require any specific electronic or nuclear structure, besides the presence of a sideband resolved transition, is applicable for alkaline-earth as well as molecular and alkali species~\cite{SM}, and does not require any auxiliary site-selective lasers, instead relying on site-selective depth control~\cite{Shaw2023A}.

\vspace{1mm}
\noindent\textbf{Motional Bell pairs and hyper-entanglement}\newline
We now turn to generating entanglement between motional states of atoms in separate tweezers; this includes preparing simultaneous Bell pairs in both the electronic and motional degrees of pairs of atoms, a state known as hyper-entanglement~\cite{Ren2014, Deng2017}.

\begin{figure*}[t!]
	\centering
	\includegraphics[width=\textwidth]{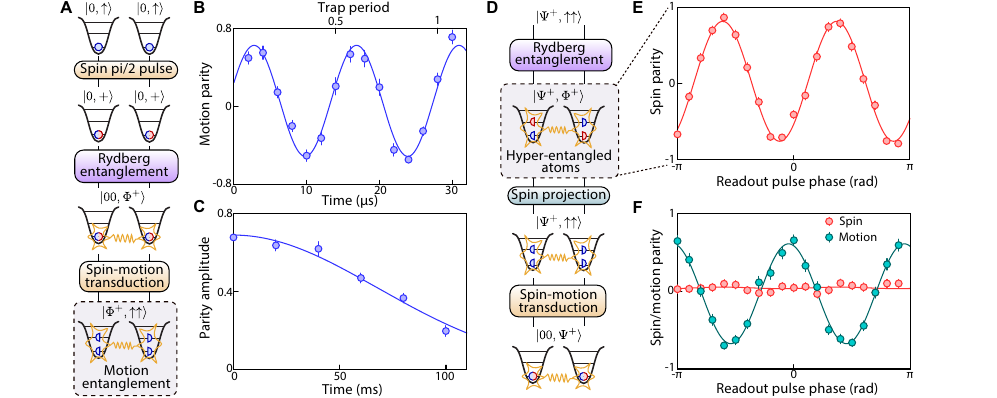}
	\caption{\textbf{Entanglement and hyper-entanglement of motion and spin.} \textbf{(A)} Protocol used to generate an entangled state of motion. After cooling the atoms and preparing the pseudo-spin superposition state $\ket{0,+}$, we entangle the spin degree of freedom via Rydberg interactions, preparing either the $\ket{00,\Phi^+}$ or the $\ket{00,\Psi^+}$ state~\cite{SM} (see text). We then apply spin-motion transduction via sideband driving, transferring spin-entanglement to motion-entanglement, yielding $\ket{\Phi^+,\uparrow\uparrow}$, or $\ket{\Psi^+,\uparrow\uparrow}$. \textbf{(B)} Parity oscillation measured by varying the time in $\ket{\Phi^+,\uparrow\uparrow}$ before converting back to pseudo-spin states for readout. The obtained oscillation frequency of $71.0^{+0.5}_{-0.5} \, \text{kHz}$ is consistent with twice the trap frequency $2\omega/(2\pi) \simeq 71 \, \text{kHz}$, as expected for $\ket{\Phi^+,\uparrow\uparrow}$. \textbf{(C)} Motion parity amplitude as a function of time. We observe a Gaussian decay of fidelity with a typical $1/e$ decay time of $96^{+5}_{-5} \, \text{ms}$. \textbf{(D)} Sketch of the protocol used to generate hyper-entanglement of both spin and motion. We prepare the motion-entangled state $\ket{\Psi^+,\uparrow\uparrow}$, and apply a second entangling operation on the spin degree of freedom, producing the hyper-Bell state $\ket{\Psi^+,\Phi^+}$. In order to read out this hyper-entanglement, we project the spin component onto $\ket{\Psi^+,\uparrow\uparrow}$, then convert the motional entanglement into spin entanglement.  \textbf{(E)}  Spin parity oscillation after hyper-entangling the atoms measured by applying a readout $\pi/2$-pulse and varying its phase. \textbf{(F)} Parity oscillation after the spin projection, without (red) and with (green) spin-motion transduction. The parity oscillation collapses without spin-motion transduction, as the projection disentangles the spin part. However, the motion entanglement remains intact, which allows us to recover spin entanglement after performing the transduction. Points represent data and lines represent fits.
 }
	\vspace{0.5cm}
	\label{Fig4}
\end{figure*}

To entangle the center-of-mass motion of two atoms, we combine spin-motion conversion and spin-spin interactions mediated via excitation to high-lying Rydberg states (Fig.~\ref{Fig4}A). We first prepare two atoms in $\ket{0,+}$, denoting the two-atom state as $\ket{00,++}$. We apply a controlled-Z gate through Rydberg interactions~\cite{Levine2019, Schine2022,Evered2023,Ma2023,SM} to prepare the spin-entangled Bell state \mbox{$\ket{00,\Phi^+} = \frac{1}{\sqrt{2}}\ket{00} \otimes (\ket{\! \downarrow \downarrow} + \ket{\! \uparrow \uparrow})$}. We then perform spin-motion transduction, obtaining a state which is spin-separable, but motion-entangled, \mbox{$\ket{\Phi^+,\uparrow\uparrow} = \frac{1}{\sqrt{2}}(\ket{00}+\ket{11})\otimes\ket{\! \uparrow\uparrow}$}. We let the state freely evolve under the trapping potential for a variable amount of time, which induces a phase difference between different motional states, followed by spin-motion transduction for readout. We observe a parity oscillation frequency of $\omega/(2\pi) \simeq 71.0^{+5}_{-5} \, \text{kHz}$, consistent with the expectation of twice the trap frequency (Fig.~\ref{Fig4}B).

Combining the contrast of the parity oscillation with the measured populations in the states $\ket{00}$ and $\ket{11}$, we obtain a motional Bell state fidelity for the state \mbox{$\ket{\Phi^+,\uparrow\uparrow}$} of $F_\text{Bell}^\text{motion} = 0.75^{+1}_{-1}$ (uncorrected for SPAM errors, but with erasure excision used to correct for errors in the preparatory spin-motion transduction~\cite{SM}). This value is limited by the bare Bell state generation fidelity in the electronic manifold (which has been improved since the data for this manuscript was taken~\cite{Finkelstein2024,Tsai2025,SM}) and the combined transduction fidelity for both atoms. We find the coherence time (fitted with Gaussian decay) of the motionally-entangled state is $96^{+5}_{-5} \, \text{ms}$ (Fig.~\ref{Fig4}C), comparable to the single-atom motional lifetime (Fig.~\ref{Fig3}G). Both the transduction fidelity and lifetime will likely improve with better trap intensity stabilization. 

These results demonstrate Bell pair generation of motional states in separate traps spaced by distances much larger than the spatial extent of the atomic wavefunction; in our particular implementation, traps are separated by $\approx3.3\ \mu$m, whereas the ground state wavefunction spread along the array axis is only ${\approx}60$ nm.

We can then go beyond pure motional entanglement, and demonstrate hyper-entanglement of motion and electronic state to realize control over an expanded Hilbert space~\cite{Deng2017}. To accomplish this task, we prepare the motion-entangled state $\ket{\Psi^+,\uparrow\uparrow}$, and then apply a second entangling operation on the electronic degree of freedom. As opposed to hyperfine or nuclear qubits, the Rydberg drive couples nearly identically to both levels of the motional qubit, up to a detuning that is much smaller than the Rabi frequency. The gate thus leaves the motional state largely unchanged, generating the hyper-Bell state $\ket{\Psi^+,\Phi^+}$.

To show that entanglement coexists on the two degrees of freedom, we measure the spin parity and population (Fig.~\ref{Fig4}E), finding an entanglement fidelity $F_\text{Bell}^\text{spin} = 0.855^{+7}_{-7}$. To confirm simultaneous motional entanglement, we perform our fast imaging procedure~\cite{Scholl2023} to project the $\ket{\Psi^+,\Phi^+}$ state onto $\ket{\Psi^+,\uparrow \uparrow}$ or $\ket{\Psi^+,\downarrow \downarrow}$ with equal probability; atoms in $\ket{{\downarrow}}$ have a low probability of surviving the fast imaging~\cite{Scholl2023}, so in the following we only consider pairs projected into $\ket{\Psi^+,\uparrow \uparrow}$. We verify that no spin entanglement remains by measuring the spin parity (Fig.~\ref{Fig4}F, red markers), finding a fitted contrast consistent with zero of $0.009^{+15}_{-15}$. We then apply spin-motion transduction, and again recover a parity oscillation signal (green markers), indicating that the motion-entanglement was preserved. We obtain $F_\text{Bell}^\text{motion} = 0.76^{+1}_{-1}$. This value is comparable to the one obtained using the sequence described in Fig.~\ref{Fig4}A, indicating that Rydberg entanglement and spin projection do not appreciably affect motion-entanglement.

\vspace{2mm}
\noindent\textbf{Outlook}\newline
This work paves the way for realizing quantum information processing tasks utilizing motional states in optical tweezers. Erasure-correction cooling can systematically outperform idealized sideband cooling, and can be applied to many other quantum platforms, including alkali and molecular species trapped in optical tweezers~\cite{SM} or optical lattices, as well as ion-based systems~\cite{Lee2023}. By coherently manipulating the resultant pure motional states, a wide array of tasks can be accomplished, such as mid-circuit readout and mid-circuit erasure detection, and including for optical qubits. Combined with entangling operations, complex hyper-entangled states over multiple internal degrees of freedom can be formed with applications in hyper-parallel quantum computation~\cite{Ren2014,Deng2017}, entanglement purification~\cite{Deng2017,Hu2021}, superdense coding~\cite{Hu2018,Graham2015}, or to measure non-trivial quantities such as Rényi entropy or fidelity overlaps via two-copy interference protocols~\cite{Pichler2016}. Given the long-lived nature of the motional entanglement, one could perform quantum computation tasks in the spin degree of freedom, and quantum memory tasks on the motional degree of freedom. This control is extensible to multiple motional axes, and to the full bosonic degree of freedom to implement quantum error correction schemes~\cite{Gottesman2001,Fluhmann2019}, or to study lattice gauge theories~\cite{Jordan2014,Macridin2022}. Control over the bosonic degree of freedom could be further extended through coherent manipulation of trap position and depth, utilizing trap anharmonicity~\cite{Grochowski2023}. Furthermore, the mid-circuit readout and motion entanglement described here could be combined to perform ancilla-based detection and quantum error correction codes~\cite{Andersen2019,Shor1995,Knill1996}. Finally, the controlled entanglement of the motion of atoms separated at mesoscopic distances could lead to entanglement-enhanced tests of short-length gravitational effects~\cite{Tino2021}.

\begin{acknowledgements}
We acknowledge insightful discussions with, and comments from, Harry Levine, Johannes Zeiher, Daniel Barredo, Arian Jadbabaie, and Xiangkai Sun. Funding: We acknowledge support from the Army Research Office MURI program (W911NF2010136), the NSF QLCI program (2016245), the Institute for Quantum Information and Matter, an NSF Physics Frontiers Center (NSF Grant PHY-1733907), the DARPA ONISQ program (W911NF2010021), the U.S. Department of Energy, Office of Science, National Quantum Information Science Research Centers, Quantum Systems Accelerator, and the NSF CAREER award (1753386). PS acknowledges support from the IQIM postdoctoral fellowship. RF acknowledges support from the Troesh postdoctoral fellowship. RBST acknowledges support from the Taiwan-Caltech Fellowship.
\end{acknowledgements}

\newpage
\clearpage
\renewcommand\theequation{S\arabic{equation}}
\setcounter{equation}{0}  
\renewcommand\thefigure{S\arabic{figure}}
\setcounter{figure}{0}  

\section*{Material and Methods}

\subsection*{Details on erasure cooling}

Here we detail the erasure correction cooling (ECC) mechanism. We start with the erasure detection of motional excitations, then describe the two methods employed to perform error correction.

\begin{figure}[ht!]
	\centering
	\includegraphics[width=\columnwidth]{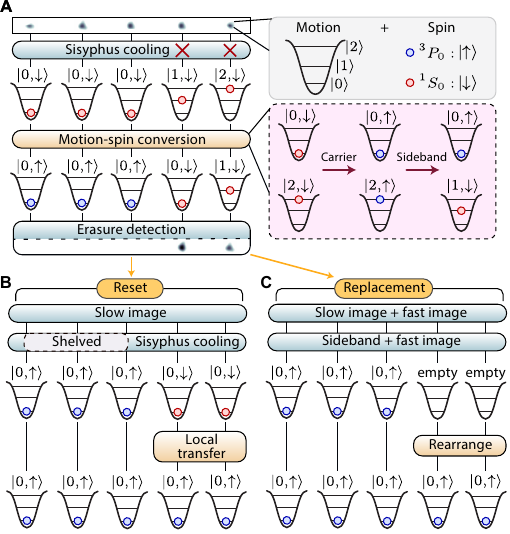}
	\caption{\label{Fig_mechanism}\textbf{Details on the erasure cooling mechanism.} \textbf{A,} After cooling the atoms using Sisyphus cooling, we transfer the atoms in $\ket{{\downarrow}}$ to the $\ket{{\uparrow}}$ state using a carrier pulse, which conserves the motional state of the atoms. We then apply a blue sideband pulse, which performs the $\ket{n,\uparrow} \rightarrow \ket{n-1,\downarrow}$ transition, where $n$ is the atom's motional level. We then selectively image atoms in $\ket{{\downarrow}}$. \textbf{B,} In the reset case, the erasure detection image is a slow and non-destructive image. After the image, we cool the atoms again. Atoms in $\ket{0,\uparrow}$ are shelved from the Sisyphus cooling mechanism. We then selectively transfer~\cite{Shaw2023B} the atoms which were imaged in $\ket{{\downarrow}}$ to $\ket{{\uparrow}}$. \textbf{C,} For the replacement method, the erasure detection image is a combination of a slow, high survival image, and a fast, low survival image. We then repeat a sideband pulse and a fast image in order to improve the erasure cooling efficiency. We then fill the emptied tweezers with auxiliary atoms which were measured to be in the $\ket{0,\uparrow}$ state.}
	\vspace{-0.5cm}
	
\end{figure}

The erasure detection is described in Fig.~\ref{Fig_mechanism}A. After cooling the atoms with Sisyphus cooling using the $^3P_1$ level, we transfer the atoms in $\ket{{\downarrow}}$ to the $\ket{{\uparrow}}$ state using a carrier pulse, which conserves the motional state of the atoms. We then apply a blue sideband pulse which performs the $\ket{n,\uparrow} \rightarrow \ket{n-1,\downarrow}$ transition, where $n$ is the atom's motional level. Atoms in $\ket{0, \uparrow}$ are not transferred back to $\ket{{\downarrow}}$. We then selectively image atoms in $\ket{{\downarrow}}$, which allows us to spatially resolve the position of motional excitations in the array. We then correct for these errors using two methods: re-using the same atoms by selectively cooling them again (reset) or replacing them with auxiliary atoms in the $\ket{0,\uparrow}$ state (replacement). Note that in the main text at the end of the correction we transfer any remaining atom in $\ket{{\downarrow}}$ to $\ket{{\uparrow}}$ using optical pumping~\cite{Madjarov2020} to avoid any vacancy defects in the array. We work with arrays of 39 total optical tweezers.

We first detail the reset method (see Fig.~\ref{Fig_mechanism}B). In this case, the erasure detection is a slow, high survival image with a fidelity of ${\approx}0.999$ and a duration of $45\,  \text{ms}$~\cite{Covey2019A}. After the image, we cool the atoms in $\ket{{\downarrow}}$ again through Sisyphus cooling, while atoms in $\ket{0,\uparrow}$ are shelved from this cooling. We then use movement-induced phase shifts to selectively transfer~\cite{Shaw2023B} to $\ket{{\uparrow}}$ the atoms which were previously imaged in $\ket{{\downarrow}}$. This process can be repeated several times in order to improve the erasure cooling efficiency.

The erasure detection image in the case of the replacement method (see Fig.~\ref{Fig_mechanism}C) is a combination of a high survival, slow image~\cite{Covey2019A} with a duration of $30\, \text{ms}$ and a fast, low survival~\cite{Scholl2023} image with a duration of $24 \, \mu s$. We combine the slow and fast images in this way because on its own the fast image has a fidelity~\cite{Scholl2023} of only ${\approx}$0.98. The slow image improves the overall imaging fidelity to ${\approx}0.999$, while the fast image serves to catch any events where an atom has decayed -- for instance due to Raman scattering events from the $\ket{{\uparrow}}$ state to the $\ket{{\downarrow}}$ state -- during the slow image. We then perform another blue sideband pulse and a fast image in order to minimize the effect of tweezer-induced heating during the erasure detection, which improves the eventual erasure cooling efficiency. Due to the low survival of the fast image, the tweezers holding atoms in $\ket{{\downarrow}}$ are empty. We refill them with auxiliary atoms which were measured to be in the $\ket{0,\uparrow}$ state. 

We ramp down the tweezers depth to $\sim 40 \, \mu \text{K}$, which corresponds to one-tenth of their nominal depth in order to reduce the effect of tweezer-induced heating such as Rayleigh scattering. In particular, we also observe that rearrangement-induced heating strongly depends on tweezers depth, and is minimized for tweezers depth in the range $\sim 20 - 40\, \mu \text{K}$. This rearrangement-induced heating is identified to be coming from (i) the finite probability that an atom is heated during transportation, and (ii) tweezers depth oscillations during atom movement due to intermodulation of the varying frequencies in the acousto-optic deflector used to generate the tweezers~\cite{Endres2016}, although we leave a quantitative analysis of these effects to future work.

\begin{figure}[t!]
	\centering
	\includegraphics[width=\columnwidth]{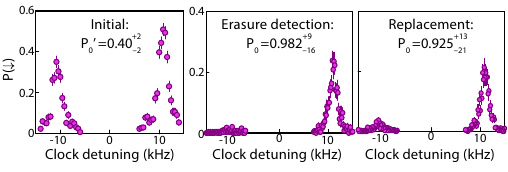}
	\caption{\textbf{Erasure correction cooling in the axial direction.} Probability $P_0$ for an atom to be in the $\ket{0}$ state measured via sideband spectroscopy, in the axial direction of the tweezers.  We show the results without erasure correction cooling, performing erasure conversion with post-selection (not correction), and with correction using the replacement method.
 }
	\vspace{0.5cm}
	\label{Fig_axial}
\end{figure}

\subsection*{Erasure cooling in the axial direction}

The erasure cooling results presented in the main text concern one of the tightly-confining radial direction. We here show that the method also applies to the axial direction, enabling this new cooling mechanism to be used to perform 3D cooling in principle. The overall method is the same as described earlier, and the only difference lies in the direction of the sideband driving beam with respect to the tweezer axis. We use a beam that propagates in the same direction as the tweezer, allowing us to drive sideband transitions on the axial direction of the tweezers, for which the trapping frequency is $10.7 \, \text{kHz}$. The results are shown in Fig~\ref{Fig_axial}. We start with a motional ground state population $P_0' = 0.40^{+2}_{-2}$ after Sisyphus cooling, which is lower than the radial direction due to the lower trapping frequency. Applying erasure correction cooling, we obtain $P_0 = 0.982^{+9}_{-18}$ after erasure detection and post-selection, and $P_0 = 0.925^{+13}_{-25}$ with active correction using the replacement method. These results are also better than what can be achieved using sideband cooling on the $^3P_1$ transition for this trapping frequency (see Fig. 3E), showing a practical advantage in using erasure correction cooling over sideband cooling.

\subsection*{Erasure correction cooling on other platforms}

\begin{figure}[t!]
	\centering
	\includegraphics[width=\columnwidth]{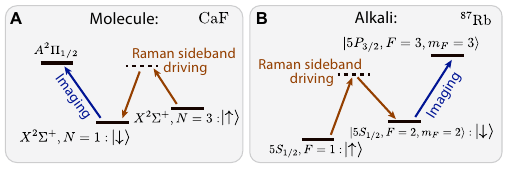}
	\caption{\textbf{Erasure correction cooling in other platforms.} \textbf{A,} Good candidates in molecules, using the example of calcium-fluoride. \textbf{B,} Good candidates in alkalis, using the example of rubidium.}
	\vspace{0.5cm}
	\label{Fig_platforms}

\end{figure}
We here show that ECC is, in principle, applicable to any tweezer platform, and detail which states can be used for the $\ket{{\downarrow}}$ and $\ket{{\uparrow}}$ when implementing ECC. Our present work uses strontium, and a generalization to other alkaline-earth(-like) atoms is trivial, so we focus on platforms using molecules and alkalis (Fig.~\ref{Fig_platforms}).

As discussed in the main text, the two key ingredients to implement ECC are (i) driving sideband transitions between $\ket{{\downarrow}}$ and $\ket{{\uparrow}}$, and (ii) the ability to selectively detect $\ket{{\downarrow}}$ while $\ket{{\uparrow}}$ is shelved. We note that the erasure cooling scheme should work even given a differential trap depth between the ground and excited electronic states (i.e. non-magic trapping) as long as an efficient sideband drive can be implemented, via e.g. a frequency chirp~\cite{Holzl2023} or a composite pulse.  These ingredients are present for alkali atoms~\cite{Kaufman2012} and also for molecules with an optical cycling transition~\cite{Anderegg2019,Vilas2024,Zhang2022B} (for which resolved sideband cooling~\cite{Lu2024,Bao2023} and molecular rearrangement~\cite{Picard2024} have recently been demonstrated). 

For molecules (Fig.~\ref{Fig_platforms}A), good candidates are two different rotational states, such as \break $\ket{{\downarrow}} = X^2\Sigma^+, N=1$ and \mbox{$\ket{{\uparrow}} = X^2\Sigma^+, N=3$} in calcium fluoride. Sideband transitions can be achieved through Raman processes, and selective imaging of $\ket{{\downarrow}}$ is achieved through driving the closed-transition to $A^2\Pi_{1/2}$, for which $\ket{{\uparrow}}$ is shelved.

For alkalis (Fig.~\ref{Fig_platforms}B), good candidates are two different hyperfine states, such as \break $\ket{{\downarrow}} = \ket{5S_{1/2},F=2,m_F=2}$ and $\ket{{\uparrow}} = \ket{5S_{1/2},F=1,m_F=1}$ in rubidium 87. Sideband transitions can be achieved through a Raman process, and selective imaging of $\ket{{\downarrow}}$ is achieved through driving the closed-transition to $\ket{5P_{3/2},F=3,m_F=3}$, for which $\ket{{\uparrow}}$ is shelved.

In principle ECC is also applicable to ion-based platforms, for which post-selection on detected erasures has been recently reported~\cite{Lee2023}. Furthermore, ECC should also work for neutral atoms and molecules trapped in optical lattices by using the reset method combined with local addressing in order to perform the local rotations~\cite{Weitenberg2011}.

\subsection*{Cooling simulations}
Here we provide details on the cooling simulations results presented in Fig.~2E. We first describe sideband cooling on the $^3P_1$ level of strontium, then the idealized sideband cooling, and finally both idealized and noisy erasure cooling using replacement.

In all cases, we simulate the following one-dimensional Hamiltonian
\begin{equation}
H = \hbar \omega(a^\dagger a +\frac{1}{2}) -\hbar\Delta\ket{{\uparrow}}\bra{\uparrow\!}+\frac{1}{2}\hbar\Omega(e^{i\eta(a+a^{\dagger})}\ket{{\uparrow}}\bra{\downarrow\!}+\text{h.c.}),
\label{eq:Ham}
\end{equation}
where $a^\dagger$ ($a$) is the creation (annihilation) operator acting on the motional levels, $\omega$ is the trapping frequency, $\Delta$ is the detuning, $\eta$ is the Lamb-Dicke factor and $\Omega$ is the Rabi frequency. We set $\Delta = -\omega$, and use a basis of ten motional states.

\subsubsection*{Ideal sideband cooling on $^3P_1$}
We perform a simulation of sideband cooling on the $\ket{^3P_1,m_J=0}$ state. Here we assume ideal conditions, namely: (i) we assume no laser noise, (ii) we set a Rabi frequency 500 times smaller than the cooling state linewidth, and (iii) we consider cooling only in a single direction, but take into account the dipole radiation pattern in 3D (more details below).

We solve the Master equation in which we take into account spontaneous emission from the $\ket{{\uparrow}}$ state and its associated momentum kick which follows a dipole pattern distribution:
\begin{equation*}
i\hbar \partial_t \rho = (H_\Gamma \rho - \rho H_\Gamma^\dagger) + i\hbar\Gamma \int d\theta D(\theta)c_{\theta}\rho c^\dagger_{\theta}.
\end{equation*}
We define $H_{\Gamma} = H-\frac{i}{2}\hbar \Gamma \ket{{\uparrow}} \bra{{\uparrow }}$, and \mbox{$c_{\theta} = e^{-i|k|\cos(\theta)}\ket{{\downarrow}}\bra{{\uparrow}}$}. Here, $\rho$ is the density matrix of the spin and motional states, $\Gamma = 2\pi \times 7.4 \, \text{kHz}$ is the linewidth of $^3P_1$, $D(\theta)=3\sin^2(\theta)/(8\pi)$ is the dipole radiation pattern for the transition of interest, $\theta$ the angle of emission with respect to the quantization axis. The dipole emission pattern takes this simple form thanks to considering cooling on the $\ket{^3P_1, m_J =0}$ state.

We use as an initial condition a thermal distribution of motional states with mean vibrational level $\bar{n} = 1$. We set $\Omega = \Gamma/500$, and perform the simulation for 1000 Rabi cycles. This ensures that we reach the system's steady state, and that the effect of the finite $\Omega$ on the final motional state populations is negligible. 

We note that in this work, we do not perform a simulation of cooling in the three directions of space, even though we take into account the dipole emission pattern in 3D. We assume that there is no thermalization between the different axes; if such cross-axis thermalization did exist it would lead to a higher temperature than we find here.

\subsubsection*{Idealized sideband cooling}
We consider sideband cooling on an idealized cooling transition, which is defined as having an infinitely narrow linewidth ($\Gamma = 0$), but still having the ability to decay to the spin ground state. We set $\Omega = 2\pi \times 2.5\, \text{kHz}$ and assume that the difference in energy between the spin states is given by a wavelength of $698 \, \text{nm}$, which would correspond to cooling on the $^3P_0$ state of strontium. We initialize the system in a thermal distribution with $\bar{n} = 1$, and perform the following simulation. We evolve the system's state for a duration corresponding to a sideband $\pi$ pulse. We then project the state onto $\ket{{\downarrow}}$ in order to mimic spontaneous decay. During this event, a momentum kick is provided to the atom following the radiation dipole pattern described earlier. We repeat this process 100 times, which ensures that we reach the system's steady state.

\subsubsection*{Idealized erasure correction cooling using replacement}

The idealized erasure correction cooling simulations are similar to the idealized sideband cooling. We set $\Omega = 2\pi \times 2.5\, \text{kHz}$ and initialize the system in a thermal distribution with $\bar{n} = 1$. We then perform the following simulation. Starting in $\ket{{\uparrow}}$, we evolve the system's state for a duration corresponding to a sideband $\pi$ pulse. In order to simulate a perfect erasure detection and replacement, we get rid of the population in $\ket{{\downarrow}}$, and re-normalize the population of the various motional levels in $\ket{{\uparrow}}$. We repeat this process 100 times, which ensures that we reach the system's steady state.

\subsubsection*{Noisy simulation of erasure correction cooling using replacement}
We finally describe the simulation of erasure correction cooling using replacement including relevant imperfections. We solve the Master equation described earlier in which we take into account the finite lifetime of the $^3P_0$ state, and the tweezer-induced heating rate, which we implement by adding a jump operator acting on the motional states. We independently measure the effect of the tweezer-induced heating, which we find increases the atomic temperature by $3.0(4)$ vibrational quanta per second. We initialize the system in a thermal distribution with $P_0'=0.77$, and simulate the sequence described in Fig.~\ref{Fig_mechanism}. We assume a perfect imaging fidelity in the erasure detection but take into account the tweezer-induced heating during the imaging and replacement times.

\subsection*{Fundamental limits of sideband cooling and erasure correction cooling}
\begin{figure}[ht!]
	\centering
	\includegraphics[width=\columnwidth]{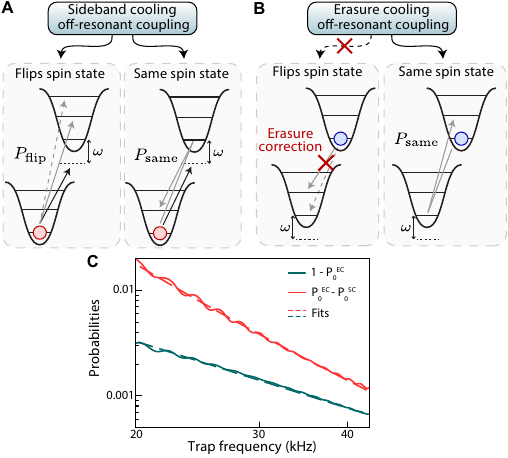}
	\caption{\textbf{Limits of sideband cooling and erasure correction cooling} \textbf{A,} The limit in sideband cooling efficiency is due to the off-resonant coupling to motional levels which are not the ground state ($n>0$), either (i) in the opposite pseudo-spin state as the initial one $\ket{n,\uparrow}$ via an off-resonant one-photon process, or (ii) in the same spin state $\ket{n,\downarrow}$ via an off-resonant two-photon process. \textbf{B,} In the case of erasure correction cooling, off-resonant coupling to the opposite spin state $\ket{n,\downarrow}$ is detected and hence does not contribute to the finite efficiency. The limit is the off-resonant two-photon transition to $\ket{n,\uparrow}$. \textbf{C,} Simulated efficiency of ideal ECC $1-P_0^\text{EC}$ (red line), and difference between sideband cooling and ECC efficiencies $P_0^\text{EC}-P_0^\text{SC}$ (green line), as a function of trap frequency $\omega$. We fit the data using $f(\omega)=A(1/\omega)^B$, with $A$ and $B$ as free parameters (dashed lines). The obtained exponents are $B=1.9^{+1}_{-1}$ for $1-P_0^\text{EC}$ and $B=2.9^{+1}_{-1}$ for $P_0^\text{EC}-P_0^\text{SC}$, which is consistent with our expectations.
 }
	\vspace{0.5cm}
	\label{Fig_fundlimits}
\end{figure}
\begin{figure*}[t!]
	\centering
	\includegraphics[]{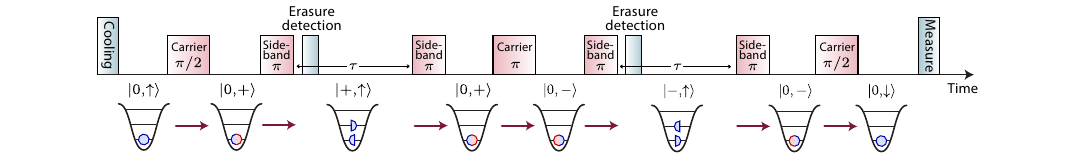}
	\caption{\textbf{Motional-echo sequence.} Sketch of the motional-echo sequence used in Fig.~\ref{Fig3}F, here assuming the phase accumulated during the wait time $\tau$ is exactly a multiple of $2\pi$ solely for presentation simplicity.
 }
	\vspace{0.5cm}
	\label{Fig_motion_echo}
\end{figure*}
In the main text, we experimentally show that erasure correction cooling (ECC) using the replacement method leads to higher $P_0$ than a simulation of idealized sideband cooling (SC), for a specific range of parameters. Here we explain why, on a fundamental level, this new cooling mechanism leads to a larger $P_0$ than SC for any trapping frequency. We further determine the scaling of the ECC efficiency in the sideband resolved and Lamb-Dicke regimes, as a function of the trap frequency. In brief, in the limit of low trap depth the sideband cooling efficiency becomes limited by (i) off-resonant coupling to higher motional states owing to the finite Rabi frequency (here $2 \pi \times 2.5\, \text{kHz}$) used to cool the atoms, and (ii) photon recoil during spontaneous emission. Conversely, ECC is (i) less sensitive to errors coming from the finite Rabi frequency because they can be partially erased, and (ii) not limited by photon recoil since cooling does not rely on spontaneous emission from a cooling state. Instead, ECC is primarily limited by tweezer-induced heating during error correction.

\subsubsection*{Fundamental limits}

We start by describing the fundamental limits of sideband cooling. We assume that the atoms are initialized in $\ket{0,\downarrow}$, and that we drive a red sideband transition in order to cool the atoms. In this situation, sideband cooling has a finite probability to actually heat the atoms via off-resonant couplings to any motional state $\ket{n,\uparrow}$. Note that in the deep Lamb-Dicke regime, spontaneous emission does not lead to a change in motional state, and thus off-resonant coupling to $\ket{0,\downarrow}$ does not impact the finite cooling efficiency.

These off-resonant couplings affect the system in two qualitatively different ways (Fig.~\ref{Fig_fundlimits}A): (i) the off-resonant one-photon coupling flips the initial pseudo-spin state to reach $\ket{n,\uparrow}$ with probability $P_\text{flip}$, and (ii) an off-resonant two-photon process which transfers the atoms in $\ket{n,\downarrow}$ and $n>0$, with probability $P_\text{same}$. The values of $P_\text{flip}$ and $P_\text{same}$ depend on the Rabi frequency, Lamb-Dicke factor, and trap frequency, and are detailed below. The eventual fundamental limit in SC is therefore the sum of these two contributions:
\begin{equation*}
    P_0^\text{SC} = 1 - P_\text{flip} - P_\text{same},
\end{equation*}
where $P_0^\text{SC}$ is the fundamental limit of the $\ket{0}$ population with sideband cooling.

We now turn to the case of ECC, and consider off-resonant coupling out of the initial state $\ket{0,\uparrow}$ (see Fig.~\ref{Fig_fundlimits}B). For ECC, off-resonant coupling to opposite spin states $\ket{n,\downarrow}$ does not impact the ECC cooling efficiency: any atom transferred to $\ket{{\downarrow}}$ is detected and replaced, which can be effectively seen as $P_\text{flip}=0$. The fundamental limit is thus given only by the pseudo-spin conserving, off-resonant two-photon transition $\ket{0,\uparrow} \rightarrow \ket{n,\uparrow}$, which cannot be converted into an erasure in our scheme. We therefore get:
\begin{equation*}
    P_0^\text{EC} = 1 - P_\text{same},
\end{equation*}
where $P_0^\text{EC}$ is the fundamental limit in $\ket{0}$ population with erasure correction cooling.

We note that the value of $P_\text{same}$ is the same in both cooling methods, as the action employed to cool the atoms -- a red sideband pulse -- is exactly the same for both methods. We therefore obtain that
\begin{equation*}
    P_0^\text{EC}-P_0^\text{SC} = P_\text{flip},
\end{equation*}
meaning that $P_0^\text{EC}$ is an upper bound for $P_0^\text{SC}$.

\subsubsection*{Scaling of $P_\text{flip}$ and $P_\text{same}$ with $\omega$}

We further comment on the scaling of $P_\text{flip}$ and $P_\text{same}$ with the trap frequency $\omega$, in the sideband resolved ($\Omega \ll \omega$) and Lamb-Dicke ($\eta(2\bar{n}+1)\ll 1$) regimes. For simplicity, we will assume in the following that the atoms are initialized in the $\ket{{\downarrow}}$ state and that we drive a red sideband transition to the $\ket{{\uparrow}}$ state. Note that the results obtained below are also true if the atoms are initialized in $\ket{{\uparrow}}$.

Under these assumptions, $P_\text{flip}$ to lowest order arises from off-resonant one-photon coupling via the $\ket{0,\! \downarrow}\rightarrow \ket{1,\! \uparrow}$ transition, with strength $\eta \Omega$ and detuning proportional to $\omega$. We therefore obtain
\begin{equation*}
    P_\text{flip} \propto \dfrac{(\eta \Omega)^2}{\omega^2} \propto \dfrac{\Omega^2}{\omega^3},
\end{equation*}
where we used $\eta \propto 1/\sqrt{\omega}$. 

$P_\text{same}$ to lowest order arises from off-resonant coupling via the $\ket{0,\! \downarrow}\rightarrow \ket{0,\! \uparrow}$ transition, with strength $(1-\eta^2) \Omega \simeq \Omega$ and detuning proportional to $\omega$. Once this off-resonant transition happens, the on-resonance red sideband transition $\ket{0,\! \uparrow}\rightarrow \ket{1,\! \downarrow}$ transfers the atoms back to the same spin state, but in a higher motional level. We therefore obtain
\begin{equation*}
    P_\text{same} \propto \dfrac{\Omega^2}{\omega^2}.
\end{equation*}

We can therefore extract the scaling of idealized sideband cooling and erasure correction cooling in the sideband resolved and deep Lamb-Dicke regimes. In particular, we obtain
\begin{equation*}
    1-P_0^\text{EC} \propto \dfrac{\Omega^2}{\omega^2},
\end{equation*}
and
\begin{equation*}
    P_0^\text{EC}-P_0^\text{SC} \propto \dfrac{\Omega^2}{\omega^3}.
\end{equation*}

We verify these scalings using our numerical simulations results of idealized SC and ECC (Fig.~\ref{Fig_fundlimits}C), which are also presented in Fig. 2F. We consider the ECC deficiency $1-P_0^\text{EC}$ (green solid line) as well as the difference between SC and ECC efficiencies $P_0^\text{EC}-P_0^\text{SC}$ (red solid line). In both cases, we fit our results with the function $f(\omega)=A(1/\omega)^B$ (dashed lines), where $A$ and $B$ are free parameters. The obtained exponents are $B=1.9^{+1}_{-1}$ for $1-P_0^\text{EC}$, and $B=2.9^{+1}_{-1}$ for $P_0^\text{EC}-P_0^\text{SC}$. These results are consistent with the formulas derived above.

\subsection*{Motional coherence and motional-echo sequence}

Here we describe the motional-echo sequence we use for the results presented in Fig. 3E. The sequence is similar to the one presented in Fig. 3A, and is further detailed in Fig.~\ref{Fig_motion_echo}. In the absence of such an echo sequence, we find relatively short motional coherence times of up to ${\sim}5 \, \text{ms}$, limited by both trap frequency inhomogeneities within the array, and global shot-to-shot fluctuations in trap frequency. Trap frequency inhomogeneities across the array are around a few percent, mostly limited by inhomogeneity in the trap waist (by contrast, trap depth inhomogeneity is only around a few tenths of a percent). Shot-to-shot global fluctuations of trap frequency are of a similar order of around a percent. While the motional echo sequence can only correct these slow variations in trap frequency, further improvements to the trap depth control feedback loop may realize gains for both slow and fast variations.

The scheme proceeds as follows: (1) we prepare the motion-superposition state $\ket{+, \! \uparrow}$ as described in the main text. (2) The state is then allowed to evolve for a duration $\tau$. (3) We then convert the state back into the electronic manifold, and apply a carrier $\pi$-pulse, which exchanges the populations in $\ket{{\uparrow}}$ and $\ket{{\downarrow}}$. (4) The state is then transduced back into the motional manifold, meaning that effectively we have exchanged the populations in $\ket{0}$ and $\ket{1}$. (5) The state then evolves for time $\tau$, realizing effectively a standard spin-echo sequence. Erasure-excision is performed on this sequence by performing a fast image just before steps (2) and (5), and post-selecting on detected atoms.

The motional-echo sequence used to obtain the results of Fig. 3E consists of four sideband pulses. In order to derive the one-way transduction fidelity $F_\text{OW}=0.993^{+3}_{-3}$ quoted in the main text from the bare measured contrast of $F_\text{tot}=0.972^{+10}_{-10}$, we assume that each sideband pulse is independent from the previous ones, meaning $F_\text{tot} = F_\text{OW}^4$. This assumption is reasonable as we perform erasure detection between sideband pulses, which both suppress any potential coherent superposition between $\ket{0,\! \downarrow}$ and $\ket{1,\! \uparrow}$, and also ejects any atom left in $\ket{{\downarrow}}$.

Finally, we note that the fundamental limit to the motional coherence time would stem from Rayleigh scattering. This scattering rate was calculated~\cite{Dorscher2018} for the fermionic $^{87}\mathrm{Sr}$ to be about $10^{-4}\, \text{s}^{-1}$ at a trap depth of a single recoil energy $U=E_r$; for comparison we typically operate at a trap depth of $U\approx250 E_r$ ($\approx1300 E_r$ during imaging).

\subsection*{Two qubit Rydberg gate}
\begin{figure}[t!]
	\centering
	\includegraphics[width=\columnwidth]{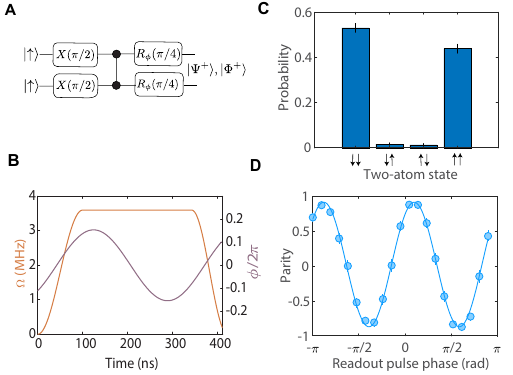}
	\caption{\textbf{Two qubit Rydberg-mediated gate.} \textbf{A,} Quantum circuit for preparation of a Bell-state in the electronic degree of freedom. \textbf{B,} The $\mathrm{CZ}$ gate is realized through single-photon excitation to a Rydberg state ($5s61s$ $^{3}S_{1}$) of two atoms spaced by 3.25 $\mu$m with a phase modulated field. \textbf{C,} Two-atom state population at the end of the circuit. We observe SPAM uncorrected $0.97^{+2}_{-2}$ population in the Bell-state $\Phi^+$. \textbf{D,} Parity oscillations induced by final $\pi/2$ pulse with variable phase. We observe a contrast of $0.89^{+1}_{-1}$ with no SPAM correction.}
	\vspace{-0.0cm}
	\label{Fig_Ry_gate}
\end{figure}

The controlled-Z ($\mathrm{CZ}$) gate is realized by a global excitation in the Rydberg blockade regime with a time-optimal pulse \cite{Jandura2022,Pagano2022,Evered2023,Ma2023}. We use a single-photon excitation to a Rydberg state ($5s61s\,^{3}S_{1}$) of two atoms spaced by 3.25 $\mu$m with a phase modulated field (Fig.~\ref{Fig_Ry_gate}). A 500G magnetic field perpendicular to the $\ket{{\downarrow}} \rightarrow \ket{{\uparrow}}$ laser beam propagation axis is applied throughout the sequence, enabling high Rabi frequency on this transition. This high magnetic field results in a reduction of the Rydberg interaction strength as compared to low field conditions~\cite{Choi2023}, and we leave an investigation of this effect to future work. We measure a $C_6$ coefficient of 52 GHz$\cdot\mu m^6$ and use a maximal Rabi frequency of $2\pi \times 3.6 \, \text{MHz}$. 

The Bell states are produced as follows. Starting with both atoms in $\ket{{\downarrow}}$, we apply a first $\pi/2$ pulse on the $\ket{{\downarrow}} \rightarrow \ket{{\uparrow}}$ transition to prepare the superposition state $\ket{+}$. We then apply the $\mathrm{CZ}$ gate, and subsequently perform a $\pi/4$ pulse to obtain a Bell state. The choice of Bell state (either $\ket{{\Psi^+}}$ or $\ket{{\Phi^+}}$) is made by changing the phase of the last $\pi/4$ pulse. For $\ket{{\Phi^+}}$, we measure SPAM uncorrected populations $P_{\uparrow\uparrow}+P_{\downarrow\downarrow}=0.97^{+2}_{-2}$; and induce parity oscillations with a final $\pi/2$ analyzing pulse where we measure a contrast of $0.89^{+1}_{-1}$, yielding a SPAM-uncorrected optical Bell state~\cite{Schine2022} generation fidelity of $0.93^{+2}_{-2}$. 

Our SPAM-uncorrected fidelity for generating the motional Bell state is roughly a combination of the fidelity for generating the optical Bell state and performing the subsequent spin-motion transduction. For this spin-motion transduction, we can apply erasure-excision as demonstrated in Fig. 3 of the main text. To readout the motional Bell state, we then need to apply another spin-motion transduction (for which we cannot perform erasure-excision), followed by a $\pi$/2 readout pulse. Multiplying the individual fidelities of these steps as reported in the main text gives a predicted fidelity of $0.80^{+2}_{-2}$, as compared to the measured value of $0.75^{+1}_{-1}$. 

We attribute the discrepancy between these two values to the fact that when performing entangling operations we operate with both a lower magnetic field and a tighter array spacing designed to favor Rydberg interactions (discussed above); however, these conditions lower our fidelity of driving the clock sideband transition and for performing erasure excision slightly from the optimum condition reported in the main text. This performance is not fundamentally limited and could be improved by widening array spacing prior to the erasure imaging, and by improving trap frequency stability and homogeneity (as traps become more inhomogeneous and difficult to control for tighter array spacings). We note that following further optimization, performed after the data for this manuscript was taken, we observe a SPAM-corrected controlled-Z fidelity of $0.9971(5)$ measured in a randomized benchmarking sequence~\cite{Finkelstein2024,Tsai2025}, with an associated Bell state generation fidelity (in the electronic manifold) of 0.989; while we have not attempted to repeat our study of motional entanglement and hyper-entanglement in this improved setting, we expect all our associated fidelities would increase accordingly.

\subsection*{Parity oscillations for the motional Bell states}

\begin{figure}[t!]
	\centering
	\includegraphics[width=\columnwidth]{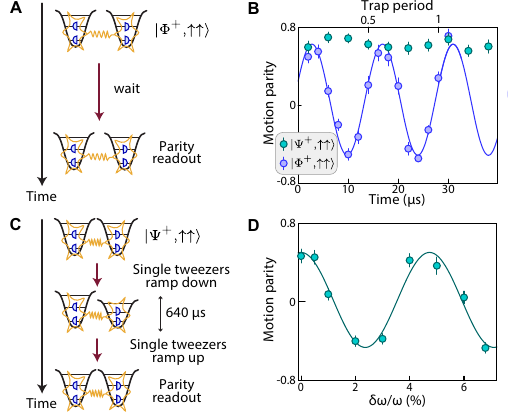}
	\caption{\textbf{Inducing parity oscillations on an odd-parity Bell state with relative trap depth variation.} \textbf{A,} Sketch of the performed experiment to induce the parity oscillation on $\ket{\Phi^+,\uparrow\uparrow}$. \textbf{B,} Parity oscillations following the experiment described in \textbf{A} for $\ket{\Phi^+,\uparrow\uparrow}$ (also shown in Fig. 4B), and for $\ket{\Psi^+,\uparrow\uparrow}$, for which we do not observe significant oscillations. \textbf{C,} Sketch of the performed experiment to induce parity oscillations on $\ket{\Psi^+,\uparrow\uparrow}$. After preparing the motion-entangled Bell state $\ket{\Psi^+,\uparrow\uparrow}$, we ramp down the power of a single tweezer within each pair for a fixed duration of $640 \, \mu\text{s}$. \textbf{D,} Motion parity as a function of the relative trap depth difference between the two tweezers.}
	\vspace{-0.0cm}
	\label{Fig_antisym}

\end{figure}
Here we show our results concerning parity oscillations of the odd-parity and even-parity motion-Bell states \mbox{$\ket{\Psi^+,\uparrow\uparrow}= (\ket{01}+\ket{10})/\sqrt{2} \otimes \ket{{\uparrow \uparrow}}$} and \mbox{$\ket{\Phi^+,\uparrow\uparrow}=(\ket{00}+\ket{11})/\sqrt{2} \otimes \ket{{\uparrow \uparrow}}$}. After preparing these states, we perform the experimental sequence described in Fig.~\ref{Fig_antisym}A, which consists of holding atoms in the motion Bell state for a variable amount of time, and then reading out the parity signal. We show the results of this experiment in Fig.~\ref{Fig_antisym}B, for both $\ket{\Phi^+,\uparrow\uparrow}$ (blue circles) and $\ket{\Psi^+,\uparrow\uparrow}$ (green circles). We observe a significant difference in their behavior: for $\ket{\Phi^+,\uparrow\uparrow}$, we obtain a parity oscillation (also shown in Fig.~4B), whereas for $\ket{\Psi^+,\uparrow\uparrow}$, we do not observe significant parity oscillations.

This behavior is expected: as both tweezers have approximately the same power and same trap frequency $\omega$, the energy spacing $\hbar \omega$ of the motional level is the same for both tweezers. 

In order to prove that we undoubtedly prepare $\ket{{\Psi^+,\uparrow\uparrow}}$, we induce parity oscillations by varying the power of a single tweezer in the pair. The experimental sequence is described in Fig.~\ref{Fig_antisym}C. After preparing the motion-entangled Bell state $\ket{\Psi^+,\uparrow\uparrow}$, we ramp down the power of a single tweezer within each pair for a fixed duration of $t_\text{wait} = 640 \, \mu\text{s}$. During this time, the state evolves as \mbox{$(\ket{01}+e^{i\delta\omega t_\text{wait}}\ket{10})/\sqrt{2} \otimes \ket{{\uparrow \uparrow}}$}, where $\delta\omega$ is the trap frequency difference between the two tweezers. 

After this step, we ramp up the tweezer power to its original value, and read-out the state's parity. We repeat this for various trap depth, and show our results as a function of $\delta\omega/\omega$ (Fig.~\ref{Fig_antisym}B). We obtain a parity oscillation, indicating that we indeed prepared the $\ket{\Psi^+,\uparrow\uparrow}$ state.

\end{document}